\documentclass[aps,prb,twocolumn,showpacs,floatfix,superscriptaddress]{revtex4-1}
\usepackage{graphicx,color}
\usepackage{amsfonts}
\usepackage[figuresright]{rotating}
\usepackage{amssymb}
\usepackage{amsmath}
\usepackage{psfrag}
\usepackage{subfigure}
\usepackage{multirow}
\usepackage{tabularx}
\usepackage{textcomp}
\usepackage{units}
\usepackage{hyperref}
\hypersetup{
 pdfnewwindow=true, colorlinks=true,
 linkcolor=blue, anchorcolor=blue,
 citecolor=blue, filecolor=blue,
 menucolor=blue, urlcolor=blue}

\usepackage{graphicx}
\usepackage{dcolumn}
\usepackage{bm}
\usepackage{color}

\makeatletter

\begin{document}
\title{Tuning the ferro- to para-electric transition temperature and dipole orientation\\of group-IV monochalcogenide monolayers}

\author{Salvador\ \surname{Barraza-Lopez}}
\email{sbarraza@uark.edu}
\affiliation{Department of Physics, University of Arkansas, Fayetteville, AR 72701, USA}
\affiliation{Institute for Nanoscale Science and Engineering, University of Arkansas, Fayetteville, AR 72701, USA}

\author{Thaneshwor P.\ \surname{Kaloni}}
\affiliation{Department of Physics, University of Arkansas, Fayetteville, AR 72701, USA}

\author{Shiva P.\ \surname{Poudel}}
\affiliation{Department of Physics, University of Arkansas, Fayetteville, AR 72701, USA}

\author{Pradeep\ \surname{Kumar}}
\affiliation{Department of Physics, University of Arkansas, Fayetteville, AR 72701, USA}

\begin{abstract}
Coordination-related, two-dimensional (2D) structural phase transitions are a fascinating and novel facet of two-dimensional materials with structural degeneracies. Nevertheless, a unified theoretical account of these transitions remains absent, and the following points are established through {\em ab-initio} molecular dynamics and 2D discrete clock models here: Group-IV monochalcogenide (GeSe, SnSe, SnTe, ...) monolayers have four degenerate structural ground states, and a 2D phase transition from a three-fold coordinated onto a five-fold coordinated structure takes place at finite temperature. On unstrained samples, the 2D phase transition requires lattice parameters to freely evolve. A fundamental energy scale permits understanding this transition. The transition temperature $T_c$ and the orientation of the in-plane intrinsic electric dipole can be controlled by moderate uniaxial tensile strain, and a modified discrete clock model describes the transition on strained samples. These results establish a general underlying theoretical background to understand structural phase transitions in 2D materials and their effects on material properties.
\end{abstract}

\date{\today}

\maketitle

\section{Introduction}\label{sec:1}

Studies of structural phase transitions in two dimensions have a long and celebrated history\cite{Potts,houtappel,Mermin} and find applications in ferromagnetism, biological and other types of membranes, polymer networks, and other soft materials.\cite{NelsonBook} Two-dimensional (2D) materials are (atom-thick) membranes too, but not much has been said concerning structural phase transitions in these materials yet.
This may be so because the most studied 2D material, graphene,\cite{2DBook,gr1,gr2} has a single (and hence non-degenerate) highly-symmetric {\em structural} ground state.

\begin{figure*}[tb]
\includegraphics[width=0.96\textwidth]{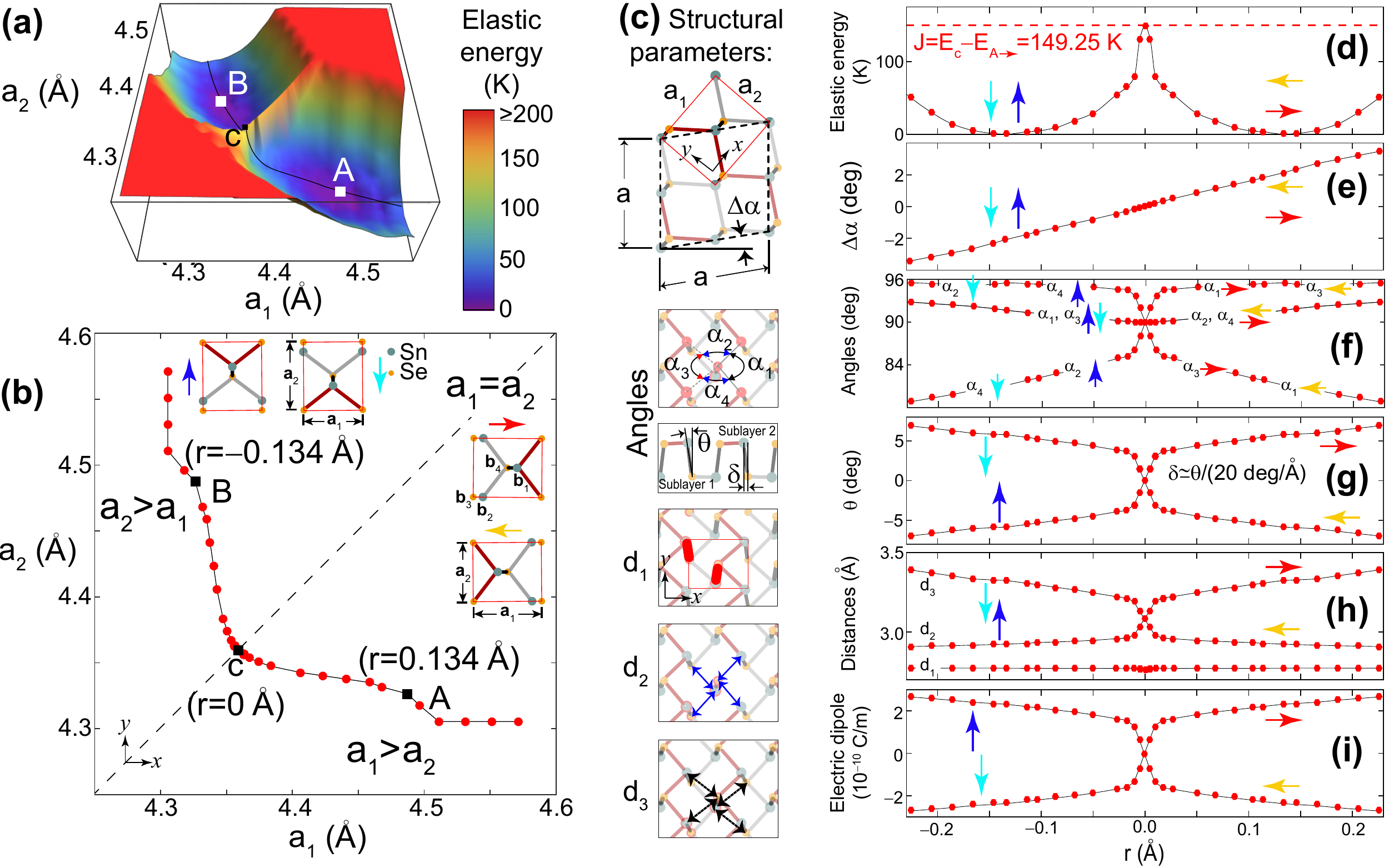}
\caption{(a) Zero-temperature energy landscape $E(a_1,a_2)$ of the unit cell of a SnSe monolayer. (b) Minimal energy pathway $E(r(a_1,a_2))$ on the landscape, joining degenerate structural ground states $A_{\rightarrow}$ and $A_{\leftarrow}$ (both located at $r=0.134$ \AA) to $B_{\uparrow}$ and $B_{\downarrow}$ (at $r=-0.134$ \AA), through the saddle point $c$. (c) Structural order parameters that signal 2D phase transitions. Zero-temperature evolution of (d) structural (elastic) energy, order parameters (e) $\Delta\alpha$, (f) angles $\alpha_1$, $\alpha_2$, $\alpha_3$, (g) $\theta$, $\delta$, (h) interatomic distances $d_1$, $d_2$, $d_3$ and (i) electric dipole as a function of $r$, for the four possible unit cells. All order parameters on subplots (d-i) depend on $r$ and therefore, on $a_1$ and $a_2$ predominantly evolving along the low-energy path drawn in (b).}\label{fig:fig1}
\end{figure*}

But graphene may rather be an exception in terms of structural degeneracies in 2D materials. Indeed, despite of its structural similarity to graphene, hexagonal boron nitride monolayers\cite{hbn} display a two-fold degeneracy by the exchange of boron and nitrogen atoms in their unit cells, silicene\cite{silicene1,silicene2,silicene3} has a two-fold degeneracy that is revealed by the exchange of upper and lower atoms in its buckled structure,\cite{Pablo,review} and transition-metal dichalcogenide monolayers in the 1T' phase (having in-plane lattice vectors that form an angle smaller than sixty degrees) are three-fold degenerate.\cite{1tprime} Unlike square\cite{Potts,Mermin} or honeycomb lattices,\cite{Fasolino} and as seen in Fig.~\ref{fig:fig1}(a), rectangular unit cells are degenerate too, by the exchange of long and short lattice constants, and display an anharmonic elastic energy profile that pushes the unit cell away from an unstable square configuration onto one out of two rectangular shapes with either $a_1>a_2$ or $a_1<a_2$. Therefore, 2D materials with rectangular unit cells such as black phosphorus (BP) and some group-IV monochalcogenide monolayers (GeSe, SnSe, SnTe, ...) are structurally degenerate as well. The initial two-fold degeneracy of the rectangular unit cell is aggravated by the disposition of basis atoms, and a reflection with respect to the axis perpendicular to the longest lattice vector yields an additional two-fold degeneracy,\cite{nl,others2} resulting in the four degenerate structural ground states shown in Fig.~\ref{fig:fig1}(b).

Previous paragraph implies that reduced structural symmetries are, in fact, a rather pervasive feature of 2D materials\cite{nl,Mehrshad,others1,others2,review} beyond graphene. These structural degeneracies are the prime ingredient for observing 2D structural phase transitions. Among other phenomena, structural degeneracies lead to non-harmonic phonon modes,\cite{Delaire} softened elastic constants, and to structural transitions that tune material properties by temperature ($T$).

BP monolayers cannot undergo 2D structural transitions and melt directly,\cite{nl} making structural degeneracies a necessary but insufficient condition for realizing 2D structural phase transitions.

Although many group-IV monochalcogenide monolayers do undergo experimentally-verified 2D structural transitions,\cite{nl,Mehrshad,KaiSCIENCE} the present understanding of these materials at finite temperature remains work in progress. A sign of the early stage of these investigations is the huge spread in theoretical estimations of the transition (Curie, critical) temperature $T_c$\cite{Mehrshad,Yang} for identical group-IV monochalcogenide monolayers, that ought to be addressed. At the same time, the thermal behavior of these two-dimensional materials provides connections among hard- and soft-condensed matter, making these results of interest to a broad audience.

To achieve a unified description of 2D structural phase transitions in group-IV monochalcogenide monolayers, the three overarching conditions for the existence of 2D structural phase transitions are enunciated in Section \ref{sec:2}. Then, the differences among the two existent theoretical models describing the ferro-to-paraelectric phase transition in group-IV monochalcogenide monolayers are indicated in Section \ref{sec:3}; one of them (called Model 1 henceforth) is based on the NPT ensemble\cite{nl,Mehrshad} (constant number of atoms, {\em pressure}, and temperature), while the other (Model 2) is based on a NVT ensemble\cite{Yang} (constant number of atoms, {\em volume}, and temperature). It is shown that the volume constraint on the latter model yields temperature-independent lattice parameters $a_1$ and $a_2$ that are inconsistent with experiment, thus leading to an overestimation of $T_c$, as can be gathered from an analysis of the relevant energy scale of these structures in Section \ref{sec:4}. In Sections \ref{sec:5} and \ref{sec:6}, the tunability of $T_c$ by uniaxial tensile strain is shown, which also permits orienting the direction of the in-plane intrinsic electric dipole after a threshold amount of strain is applied. Section \ref{sec:7} showcases a two-parameter model that describes all observed details of these transitions qualitatively. The results provided here unify what are at the moment conflicting theoretical accounts of these structural transitions.\cite{Mehrshad,Yang} Conclusions are provided afterwards.

Considering readability for a wide audience, a deliberate effort is made to highlight physical behavior over numerics, so that descriptions of computational methods appear at the end. Although the material chosen here is SnSe, the results here are meant to describe the general behavior of group-IV monochalcogenides with rectangular unit cells.

\section{Conditions for the occurrence of 2D structural phase transitions}\label{sec:2}

To create 2D structural phase transitions, the degeneracies indicated in previous Section must be complemented by two additional conditions that are illustrated on a SnSe monolayer next:
\begin{enumerate}
\item{}In the elastic energy landscape\cite{WalesBook} $E(a_1,a_2)$ shown in Fig.~\ref{fig:fig1}(a), an energy pathway must exist that is highlighted as $r(a_1,a_2)$ in Fig.~\ref{fig:fig1}(b) and joins pairs of degenerate ground states. The joining paths are labeled $A_{\rightarrow} \leftrightarrow B_{\uparrow}$,  $A_{\rightarrow} \leftrightarrow B_{\downarrow}$, $A_{\leftarrow} \leftrightarrow B_{\uparrow}$, or $A_{\leftarrow} \leftrightarrow B_{\downarrow}$ and proceed against an energy barrier $J\equiv(E_c-E_{A_{\rightarrow}})<k_BT_m$ at point $c$, where $T_m$ is the material's melting point, $k_B$ is Boltzmann's constant, and $E_c$ is the smallest structural energy along the $a_1=a_2$ line ($E_c=min\{E(a_1,a_1)\}$) on a structure lacking electric polarization (hence the omission of arrows on $E_c$). Horizontal (vertical) arrows indicate a net dipole moment along the $x-$ ($y-$)direction.\cite{nl} Unit cells switch among any of the {\em four} degenerate structures once the barrier $J$ is overcome.\cite{Potts,nl}
\item{}Thermodynamic equilibrium requires degenerate ground states to be evenly sampled, and this implies that macroscopic domains representing the four degenerate ground states will be visible on a sample. Therefore, the second condition is that sufficiently large domains exist below $T_c$. This condition is verified by experiment.\cite{KaiSCIENCE}
\end{enumerate}
When structural degeneracies exist and conditions (1-2) are satisfied, 2D structural phase transitions alter the properties of 2D materials in ways that are only beginning to be studied.\cite{others2,nl,Mehrshad,Yang}

As displayed in Fig.~\ref{fig:fig1}(c), $\Delta\alpha$ is a geometrical variable motivated by experiment\cite{KaiSCIENCE} that signals a departure from a square unit cell ($\Delta \alpha=0$ and $a_1=a_2$) onto a rhombus ($\Delta \alpha\ne 0$ and $a_1\ne a_2$). In Fig.~\ref{fig:fig1}(c), the long and short diagonals of the rhombus are orthogonal, and have magnitudes $2a_1$ and $2a_2$, respectively.

Experimentally, the 2D structural transition was linked to a sudden collapse of $\Delta\alpha$ to zero\cite{KaiSCIENCE} which, in turn, requires a sudden change of lattice parameters at the Curie temperature $T_c$ onto $a_1/a_2=1$,\cite{nl,Mehrshad} (see Fig.~\ref{fig:fig1}(c)):
\begin{equation}\label{eq:1}
\frac{a_1(r)}{a_2(r)}=\frac{1+\sin\Delta \alpha(r)}{\cos\Delta \alpha(r)}(\simeq 1+\Delta\alpha(r) \text{ for }\Delta\alpha(r)\simeq 0).
\end{equation}
According to Eqn.~\eqref{eq:1}, $\Delta\alpha= 0$\cite{KaiSCIENCE} implies $a_1=a_2$\cite{nl,Mehrshad} and $r=0$ in Fig.~\ref{fig:fig1}(b). The reader must note that no other theory exists at this moment that reproduces this experimental fact.

Besides $\Delta\alpha$, the four basis atoms ($\mathbf{b}_i$, $i=1,2,3,4$) confer this 2D material with additional structural order parameters: distances $d_1=|\mathbf{b}_3-\mathbf{b}_2|$, $d_2=|\mathbf{b}_4-\mathbf{b}_2|$, and $d_3=|\mathbf{b}_4-\mathbf{b}_2+\mathbf{a}_1|$; angles $\alpha_1=\angle(\mathbf{b}_2+\mathbf{a}_1,\mathbf{b}_4,\mathbf{b}_2+\mathbf{a}_1+\mathbf{a}_2)$, $\alpha_2=\angle(\mathbf{b}_2+\mathbf{a}_1+\mathbf{a}_2,\mathbf{b}_4,\mathbf{b}_2+\mathbf{a}_2)$,
$\alpha_3=\angle(\mathbf{b}_2+\mathbf{a}_2,\mathbf{b}_4,\mathbf{b}_2)$ and
$\alpha_4=\angle(\mathbf{b}_2,\mathbf{b}_4,\mathbf{b}_2+\mathbf{a}_1)$; the angle $\theta=\text{acos}\left[(\mathbf{b}_4-\mathbf{b}_1)\cdot \hat{z}\right/|\mathbf{b}_4-\mathbf{b}_1|]$,\cite{Yang} with $\hat{z}=(0,0,1)$; and $\delta=b_{1x}-b_{4x}$, the projection of the $\mathbf{b}_{1}-\mathbf{b}_{4}$ vector onto the $x-$axis.

Considering structure $A_{\rightarrow}$ for reference, the interdependence of $\delta$ and $\theta$ on $a_1(r)$, $d_1(r)$, and $\alpha_1(r)$ in Fig.~\ref{fig:fig1} is as follows:
\begin{equation}
\delta=\frac{a_1}{2}-d_2\cos\left(\frac{\alpha_1}{2}\right)\text{, and }\theta=\text{arcsin}\left(\frac{\delta}{d_1}\right).
\end{equation}
In order for the dipole moment to point along the positive $x-$direction, the chalcogen atom (1 and 3) has an $x-$coordinate smaller than the $x-$coordinate of the group-IV atom ($b_{1x}<b_{3x}$, and $b_{2x}<b_{4x}$).

$E(r)$ in Fig.~\ref{fig:fig1}(d) is a one-dimensional cut of the elastic energy landscape, Fig.~\ref{fig:fig1}(a), along the minumum energy line $r(a_1,a_2)$ displayed as Fig.~\ref{fig:fig1}(b), that emphasizes the four degenerate ground states ($A_{\rightarrow}$, $A_{\leftarrow}$, $B_{\uparrow}$ and $B_{\downarrow}$). This energy profile has a direct dependence on $a_1$ and $a_2$, as it requires both lattice parameters to vary. Negative values of $r$ in Fig.~\ref{fig:fig1}(b) --occurring for values of $a_1$ and $a_2$ such that $a_2 > a_1$-- correspond to structures with an electric dipole oriented along the vertical direction, while positive values of $r$ --taking place when $a_2< a_1$-- describe structures with a horizontal electric dipole. $E(r)$ displays a cusp at $r=0$ (point $c$ in Fig.~\ref{fig:fig1}(b)), representing a square structure with a zero net electric dipole. The existence of two minima points $A$ and $B$ in Figs.~\ref{fig:fig1}(a) and \ref{fig:fig1}(b)) implies that the elastic energy profile is anharmonic.\cite{Mao_2015}

When discussing the stability of 2D materials, Fasolino, Loss and Katsnelson argue that anharmonic contributions to the elastic energy, that are absent in the Mermin-Wagner theorem,\cite{Mermin} are crucial to understand long-range order in 2D materials. The anharmonic contribution in graphene is due to the coupling of in-plane (stretching) and out-of-plane (bending) vibrational modes.\cite{Fasolino} As shown in Fig.~\ref{fig:fig1}(d), group-IV monochalcogenide monolayers have an anharmonic elastic profile even without considering out-of-plane bending, that may render Mermin-Wagner theorem unapplicable as well.

In Figs.~\ref{fig:fig1}(e) to \ref{fig:fig1}(h), the dependence of order parameters on $r$ are shown for the four possible structures that were labeled with colored arrows, while Fig.~\ref{fig:fig1}(i) displays the dependence of the in-plane electric dipole.

Numerical details aside, the points from Fig.~\ref{fig:fig1} are as follows: (a) there are four degenerate ground states on group-IV monochalcogenide monolayers and (b) a single characteristic energy barrier $J$ to describe this 2D transition;\cite{nl} (c) the 2D transition is driven by a sudden collapse of $a_1/a_2$ to unity.\cite{nl,Mehrshad}

\begin{figure}[tb]
\includegraphics[width=0.47\textwidth]{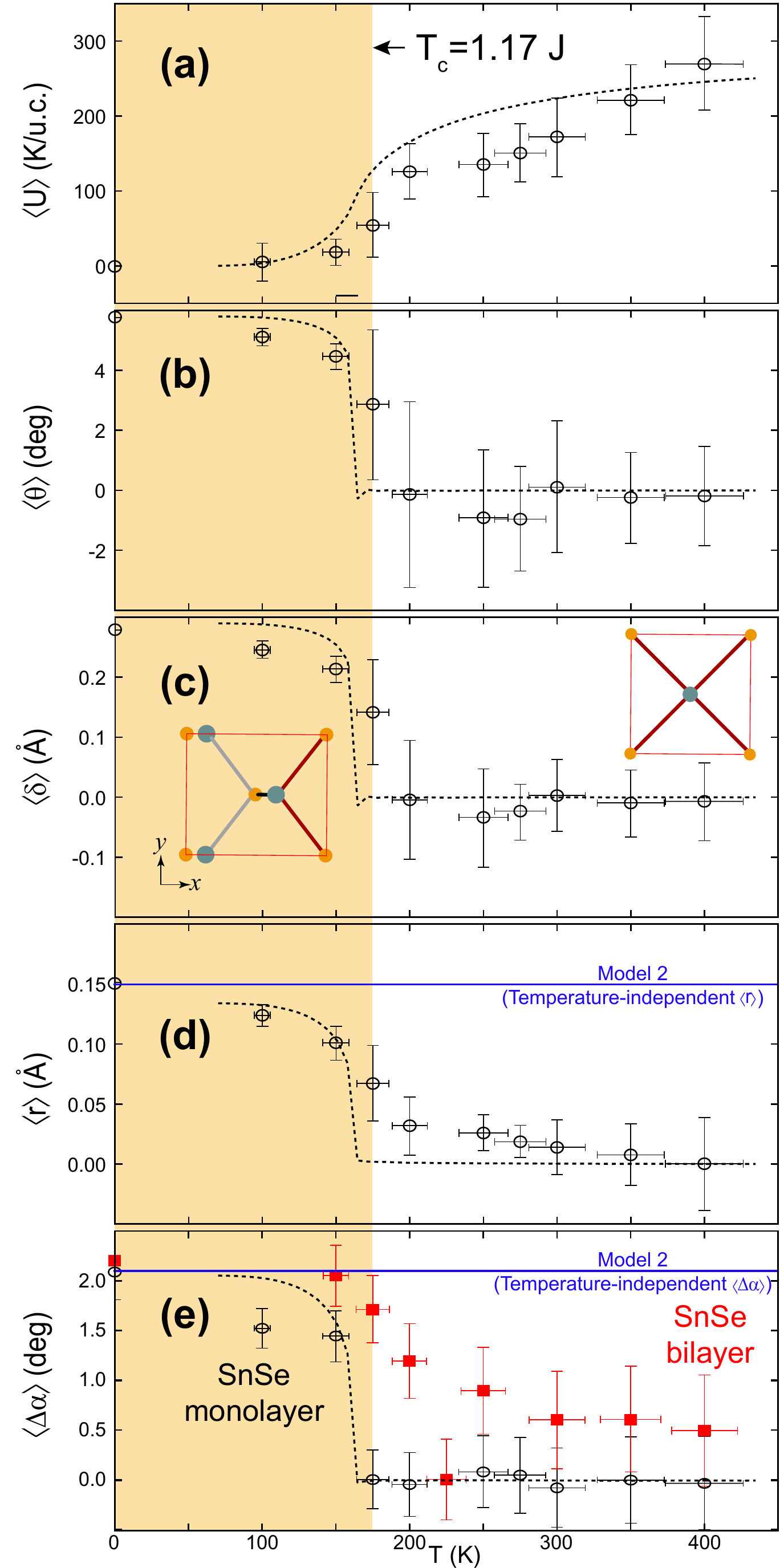}
\caption{Thermal evolution of (a) configuration energy $\langle U\rangle$, (b) $\langle \theta \rangle$,
(c) $\langle \delta \rangle$, (d) $\langle r \rangle$, and
(e) $\Delta\alpha$ for SnSe monolayers. $\langle \rangle$ stands for thermal averages. $\langle \theta \rangle$,
$\langle \delta \rangle$, $\langle r \rangle$, and
 $\langle\Delta\alpha\rangle$ all turn to zero near $T_c$.  $\langle\Delta\alpha\rangle$ is also shown for a SnSe bilayer, which displays a larger $T_c$. Straight lines in (d) and (e) display the independence of $\langle r \rangle$ and $\langle\Delta\alpha\rangle$ on temperature in the NVT ensemble. Fits originate from a discrete clock model.}\label{fig:fig2}
\end{figure}

\section{NPT ensemble and the ferro- to para-electric transition}\label{sec:3}

The zero-temperature evolution of order parameters as a function of $r$ in Figs.~\ref{fig:fig1}(e) to \ref{fig:fig1}(i) provides insight into the structural properties of this material family at finite temperature as long as $r$ (and hence $a_1$ and $a_2$) varies with $T$: molecular dynamics (MD) calculations at finite temperature carried out within the NPT ensemble (constant number of particles, constant pressure, and constant temperature) allow the lattice parameters and hence $r$ to adapt with $T$.\cite{nl} In fact, allowing $a_1$ and $a_2$ to vary is standard practice in studies of 2D materials at finite temperature.\cite{Fasolino}

One employs condition 2 from Sec.~1 and builds a simulation $8\times 8$ supercell with atoms on the ground state $A_{\rightarrow}$ configuration; {\em i.e.}, domain $A_{\rightarrow}$ is set as the initial structure at zero Kelvin. From now on, parameters within angular brackets represent thermal averages.

The structural contribution $\langle U\rangle$ to the total energy in the MD calculation is displayed in Fig.~\ref{fig:fig2}(a), showing a sudden increase at $T_c$ which implies, by virtue of Figs.~\ref{fig:fig1}(a) and \ref{fig:fig1}(d), a transition onto a square structure.

Indeed, starting on a structure originally consistent with the $A_{\rightarrow}$ structural ground state, Fig.~\ref{fig:fig2}(a) demonstrates that temperature drives the structural energy $\langle U\rangle$ up, making all other three structures ($B_{\uparrow}$, $B_{\downarrow}$, and accordingly $A_{\leftarrow}$) accessible, and thus driving the 2D structural transition. $\langle U\rangle$ is listed per unit cell in order to write it in units of temperature, which is an intensive quantity. The (yellow) box in Fig.~\ref{fig:fig2} highlights the magnitude of $T_c$ obtained in MD calculations of SnSe monolayers without uniaxial strain.\cite{Mehrshad} The trendlines are the result from Potts model, which takes $J$ as its only (fitting) parameter, and whose methodology will be described later on.

Figures~\ref{fig:fig2}(b) and \ref{fig:fig2}(c) continue to indicate that one can understand the finite-temperature behavior of these 2D materials through an assessment of their structural degeneracies and the single energy barrier $J$ at zero temperature:\cite{nl,Mehrshad} structural variables $\theta$ and $\delta$ in Fig.~\ref{fig:fig1}(c) and Eqn.~\eqref{eq:2} turn the in-plane electric dipole off at point $c$ ($r=0$), which represents a square unit cell. In a similar fashion, $\langle \theta \rangle$ and $\langle \delta \rangle$ in Figs.~\ref{fig:fig2}(b) and \ref{fig:fig2}(c)  correlate with the vanishing of $\langle r\rangle$ in Fig.~\ref{fig:fig2}(d) when thermally driven on MD runs. The larger path in Fig.~\ref{fig:fig2}(d) when contrasted with the value of $r$ at point $A$ in the zero-temperature plot (Fig.~\ref{fig:fig1}(b)) has to do with a thermal expansion of the unit cell at finite temperature.

Using Eqn.~\eqref{eq:1}, $\langle a_1 \rangle$ and $\langle a_2\rangle$ obtained from MD runs for SnSe monolayers and bilayers are recast onto the $\langle\Delta\alpha\rangle$ {\em versus} temperature plot in Fig.~\ref{fig:fig2}(e). The evolution of $\langle\Delta\alpha\rangle$ on few-layer SnTe  in Ref.~\cite{KaiSCIENCE} leads to a Curie's temperature $T_c$ that is determined by (i) a sudden collapse of $\langle\Delta\alpha\rangle$ to zero, and (ii) $T_c$ increases with the number of layers. Experimental features (i) and (ii) are generic to few-layer monochalcogenides, and captured in Fig.~\ref{fig:fig2}(e) for SnSe. This structural transition takes place within 0.8 ps in our MD calculations, an ultra-fast time that is even consistent with experimental switching times on ultrathin chalcogen-based materials.\cite{Lindenberg}

A phenomenological order-disorder model for two-dimensional phase transitions in two-dimensional crystals with four nearest-neighbors interactions and four degenerate ground states that is consistent with MD data was developed some time ago.\cite{Potts} Drawing an analogy between in-plane electric dipoles pointing along four discrete orientations and spins, $T_c$ can be estimated from a discrete 2D clock model to be:\cite{Potts,nl}
\begin{equation}\label{eq:2}
T_c=1.1(4) J/k_B.
\end{equation}
(Potts writes $\frac{2J}{k_BT_c}=1.76$ for the $r=4$ structural ground states in the present problem.)

The ratio among the magnitude of $T_c=175\pm 11 K$ --obtained through dedicated MD runs-- and $J$ --computed on a single unit cell calculation at zero temperature-- yields $T_c=1.1(7)J/k_B$, right on target with the classic work by Potts. This agreement validates the MD methodology against a classical model for phase transitions in two-dimensional lattices that is based on the single, physically-motivated parameter $J$.

In contrast, the thermal behavior of a structure with temperature-independent lattice parameters, Model 2,\cite{Yang,Rangel} is described by the NVT ensemble, where the area of the 2D material is kept fixed during the thermal evolution. Independency of lattice parameters with temperature yields $\partial (\langle\frac{a_1}{a_2}\rangle)/\partial T=0$ and, using Eqn.~\eqref{eq:1}:
\begin{equation}
\partial\langle\Delta\alpha\rangle/\partial T=0,
\end{equation}
which is inconsistent with experimental observation.\cite{KaiSCIENCE} In Figs.~\ref{fig:fig2}(d-e), $r$ and $\Delta\alpha$ in Model 2 take  constant, temperature-independent values that are emphasized by straight (blue) lines $a_i(T)=a_i(T=0)$ for $i=1,2$.

In addition to a temperature-independent $\langle\Delta\alpha\rangle$, $T_c$ is overestimated in Model 2, in the sense that Eqn.~\eqref{eq:2} is not satisfied either: working with SnSe as a representative example, the value of $T_c$ obtained in {\em ab initio} MD calculations in Model 1 is 175 K,\cite{Mehrshad} but 326 K in Model 2.\cite{Yang} Such discrepancy may hamper further work on the area, as both estimations were made with the same underlying numerical approach (pseudopotential-based density functional theory), and deserves careful attention.

The discrepancy on $T_c$ is resolved by reaching an agreement on the intrinsic energy scale that triggers the structural transition. This appears necessary, as even reported values of $a_1$ and $a_2$ display a large scatter of 4.35--4.70 and 4.24--4.40, respectively\cite{PhysRevB.92.085406,YangAPL,Lijcp,Hennig-apl,scirep,PhysRevLett.118.227401,Wang-JAP} that affects estimates of $J$ directly, and of $T_c$ through Eqn.~\eqref{eq:2}.

\begin{table}[tb]
 \caption{Optimal magnitudes for lattice and basis vectors listed in Eqns.~\eqref{eq:5} through Eqn.~\eqref{eq:8}, and energy barrier $J$ of SnSe monolayers, as obtained with three commonly used computational tools; van der Waals corrections are included in these estimates. Subindex $c$ refers to the square structure at point $c$, while subindex $A$ is to label the structural ground state $A$; c.f., Figs.~\ref{fig:fig1}(a)-(b). $z_{2c}$ and $z_{2A}$ are 0 \AA{} throughout.}\label{ta:ta1}
\centering
\begin{tabular}{|c|}
\hline
\hline
VASP, vdW.\cite{VASP,pawvasp,vaspvdw1,vaspvdw2} $(E_c-E_{A_{\rightarrow}})/k_B=154.84$ K\\
\hline
$a_c=4.3418$, $z_{1c}=2.8129$, $z_{3c}=2.7347$, $z_{4c}=0.0781$\\
$a_{1A}=4.4678$, $a_{2A}=4.2957$\\
$\delta=0.2879$, $z_{1A}=2.8641$, $z_{3A}=2.7309$, $z_{4A}=0.1331$\\
$a_{1A}/a_{2A}=1.0401$, $a_{1A}/a_C=1.0290$\\
\hline
\hline
Quantum Espresso, vdW.\cite{QE,paw,QEvdw0,QEvdw1,QEvdw2,QEvdw3} $(E_c-E_{A_{\rightarrow}})/k_B=146.04$ K\\
\hline
$a_c=4.3137$, $z_{1c}=2.8422$, $z_{3c}=2.7202$, $z_{4c}=0.1220$\\
$a_{1A}=4.4251$, $a_{2A}=4.2690$\\
$\delta=0.2684$, $z_{1A}=2.8593$, $z_{3A}=2.7180$, $z_{4A}=0.1417$\\
$a_{1A}/a_{2A}=1.0366$, $a_{1A}/a_C=1.0258$\\
\hline
\hline
SIESTA, vdW.\cite{siesta,soler,BH,pseudospaper} $(E_c-E_{A_{\rightarrow}})/k_B=149.26$ K\\
\hline
$a_c=4.3590$, $z_{1c}=2.7661$, $z_{3c}=2.7616$, $z_{4c}=0.0042$\\
$a_{1A}=4.4873$, $a_{2A}=4.3264$\\
$\delta=0.2785$, $z_{1A}=2.8035$, $z_{3A}=2.7578$, $z_{4A}=0.0457$\\
$a_{1A}/a_{2A}=1.0372$, $a_{1A}/a_C=1.0294$\\
\hline
\hline
\end{tabular}
\end{table}

\section{Intrinsic energy scale for phase transitions on 2D materials with structural degeneracies}\label{sec:4}

The intrinsic energy scale on materials with structural degeneracies is given by the energy difference among the (degenerate) ground state unit cell, and the unit cell with high symmetry at zero temperature.

The discovery of ferroelectricity on SnTe monolayers implies that this material hosts a rectangular unit cell,\cite{KaiSCIENCE} while mean-field structural calculations with DFT and the PBE approximation indicate the unit cell to be square.\cite{Yang} Using SnSe as a  representative case example, it will be shown that van der Waals corrections help increase the anisotropy among $a_1$ and $a_2$ on group-IV monochalcogenides, {\em even at the monolayer limit},\cite{Mehrshad} and may lead to structural estimates that are closer to experiment.

As indicated by condition 1 in Section \ref{sec:2}, the energy difference among the square unit cell ($E_c$) and the energy for a structure in the ground state ($E_{A_{\rightarrow}}$) yields $J$, which will be estimated in a detailed manner next.

\begin{table}[tb]
\caption{Lattice and basis vectors for structures employed to obtain SnSe energy barrier using identical methods and computational tool as in Ref.~\cite{Yang}. $b_{2zc}$ and $b_{2zc}$ are 0 \AA{} throughout. Note that lack of van der Waals corrections yields a ratio $a_{1A}/a_{2A}$ smaller than the one listed in Table \ref{ta:ta1}.}\label{ta:ta2}
\centering
\begin{tabular}{|c|}
\hline
\hline
VASP, PBE.\cite{VASP,pawvasp,PBE} $(E_c-E_{A_{\rightarrow}})/k_B=50.30$ K\\
\hline
$a_c=4.3179$, $z_{1c}=2.7256$, $z_{3c}=2.7180$, $z_{4c}=0.0077$\\

$a_{1A}=4.3819$, $a_{2A}=4.2940$\\
$\delta=0.2106$, $z_{1A}=2.7505$, $z_{3A}=2.7167$, $z_{4A}=0.0338$\\
$a_{1A}/a_{2A}=1.0205$, $a_{1A}/a_C=1.0148$\\
\hline
\hline
\end{tabular}
\end{table}

Lattice vectors for the (square) unit cell at point $c$ in Fig.~\ref{fig:fig1}(b) are given by:
\begin{eqnarray}\label{eq:5}
\mathbf{a}_1=&(a_c,0,0)\text{, }\mathbf{a}_2=(0,a_c,0)\text{, }\nonumber\\
\mathbf{a}_3=&(0,0,20\text{ \AA}),
\end{eqnarray}
while the basis vectors (that yield a zero net electric dipole given that $\theta=0$) are:
\begin{eqnarray}\label{eq:6}
\mathbf{b}_1=&(a_c/2,a_c/2,z_{1c})\text{ (Sn), }\nonumber\\
\mathbf{b}_2=&(0,0,0)\text{ (Sn),}\nonumber\\
\mathbf{b}_3=&(0,0,z_{3c})\text{ (Se), }\nonumber\\
\mathbf{b}_4=&(a_c/2,a_c/2,z_{4c})\text{ (Se),}
\end{eqnarray}
where the atomic species are indicated. $a_c$, $z_{2c}$, $z_{3c}$, and $z_{4c}$, as obtained with van der Waals corrections appear in Table \ref{ta:ta1} (numerical details are given in Section \ref{sec:methods}).

The magnitude of $a_c$ in Table \ref{ta:ta1} renders the minimal energy of a unit cell under the constraint $a_1=a_2$ on a structure that lacks an in-plane electric dipole (Eqn.~\ref{eq:6}), as necessary for all four dipole orientations to occur with equal probability as soon as $a_1\ne a_2$. Point $c$ is a saddle point on the elastic energy landscape $E(a_1,a_2)$ in Fig.~\ref{fig:fig1}(a): a minimum along the $a_1=a_2$ line, and a maximum along the (orthogonal) $r-$line in Fig.~\ref{fig:fig1}(d).

The ground state structures $A_{\rightarrow}$ and $A_{\leftarrow}$ have the following lattice vectors:
\begin{eqnarray}\label{eq:7}
\mathbf{a}_1=&(a_{1A},0,0)\text{, }\mathbf{a}_2=(0,a_{2A},0)\text{, }\nonumber\\
\mathbf{a}_3=&(0,0,20\text{ \AA}).
\end{eqnarray}
The values of $a_{1A}$ and $a_{2A}$ in Tables \ref{ta:ta1} and \ref{ta:ta2} are guaranteed to yield the minimum energy by an explicit meshing procedure for $a_1$ and $a_2$ around point $A$ that explicitly shows higher structural energies for values of $a_1$ and $a_2$ in the closest vicinity of the listed $a_{1A}$ and $a_{2A}$, that can thus be considered reliable (DFT-vdW) mean field values. The basis vectors of a ground state structure are:
\begin{eqnarray}\label{eq:8}
\mathbf{b}_1=&(a_{1A}/2\pm \delta,a_{2A}/2,z_{1A})\text{ (Sn), }\nonumber\\
\mathbf{b}_2=&(\pm \delta,0,0)\text{ (Sn),}\nonumber\\
\mathbf{b}_3=&(0,0,z_{3A})\text{ (Se), }\nonumber\\
\mathbf{b}_4=&(a_{1A}/2,a_{2A}/2,z_{4A})\text{ (Se),}
\end{eqnarray}
where a positive (negative) sign renders structure $A_{\rightarrow}$ ($A_{\leftarrow}$) that has an in-plane dipole moment that is oriented towards the positive (negative) $x-$axis, as confirmed by Bader charge analysis and Berry-phase calculations. Exchange of $x-$ and $y-$ components on both lattice and basis vectors renders the two additional degenerate structures $B_{\uparrow}$ and $B_{\downarrow}$.

The energy barrier obtained for a SnSe monolayer in Table \ref{ta:ta2} follows the exact methodology and the numerical code listed in Model 2. (They indicate that no van der Waals corrections were included in monolayer calculations.) The value $J=50.3$ $K$ for the SnSe monolayer is similar to the previously reported value,\cite{Yang} and smaller to the magnitude of 146.0--154.8 obtained with van der Waals corrections in Table \ref{ta:ta1}.

This way, the Curie temperature of 326 $K$ for a SnSe monolayer reported in Ref.~\cite{Yang} disagrees with the classic theoretical result, $T_c=1.14\times 50.3$ K, Eqn.~\eqref{eq:2}, by about 600\%. Noting that the constrained Model 2 has only two degenerate states instead of four, Potts prediction will turn into the prediction for an Ising system (i.e., the relation among $T_c$ and $J$ on a square lattice with two-degenerate structural ground states): $k_BT_c=2.27J$, which still remains 212 K below the value reported by Fei and coworkers. (In looking for a close correspondence, one should not turn inconsistent and use $J$ from a calculation with van der Waals corrections on an estimate of $T_c$ obtained with a PBE exchange-correlation potential.)

An explanation for the large value of $T_c$ in Ref.~\cite{Yang} will next be provided to solve contradicting accounts for the transition temperature, thus contributing to an unified framework to understand 2D structural transitions in these materials.

\begin{figure}[tb]
\includegraphics[width=0.45\textwidth]{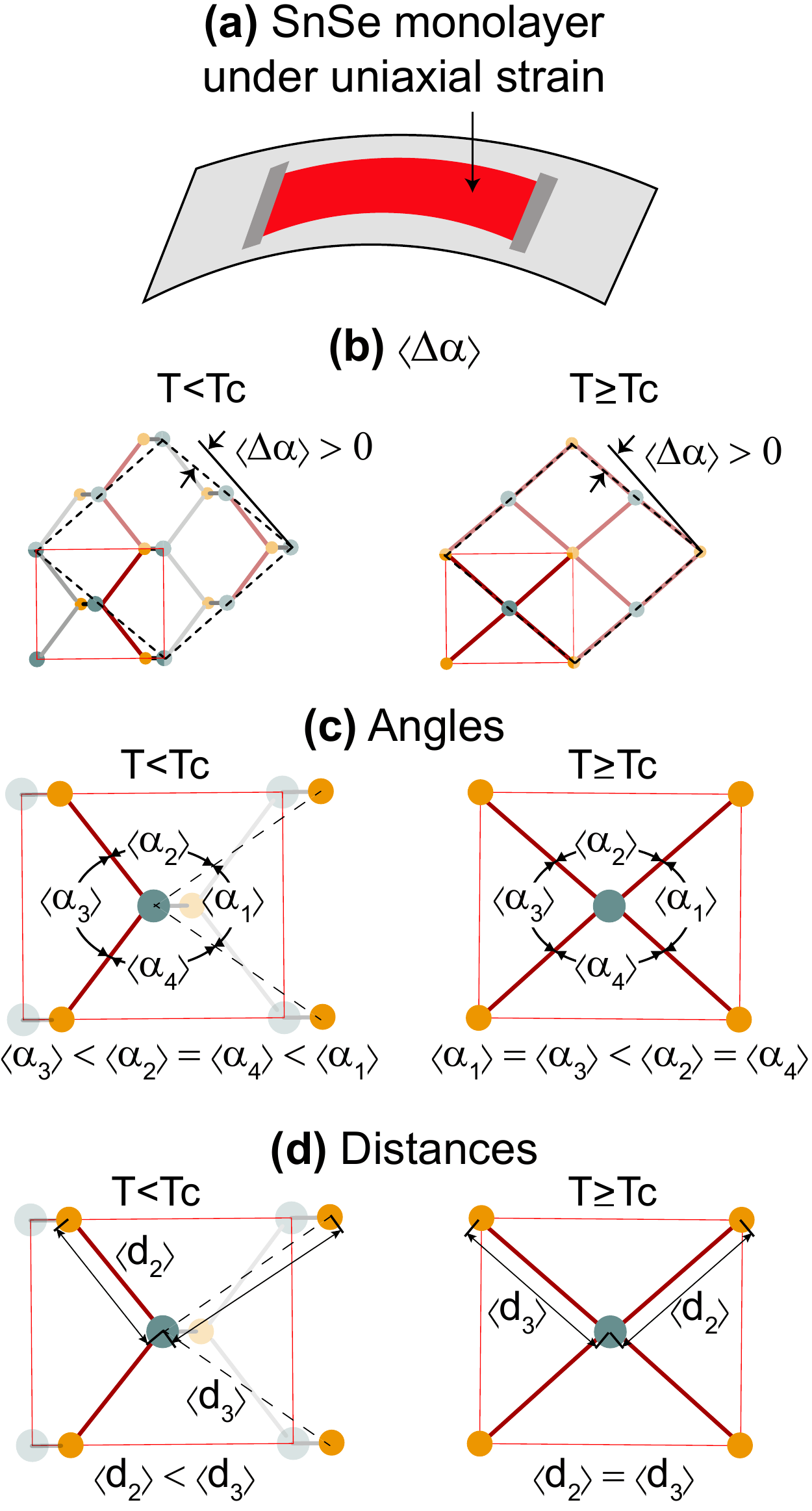}
\caption{(a) SnSe monolayer clamped onto a substrate and subjected to uniaxial tensile strain by bending. (b) to (c): proposed thermal evolution of $\langle\Delta \alpha\rangle$, $\langle \alpha_1\rangle$, $\langle \alpha_2\rangle$, $\langle \alpha_3\rangle$, $\langle \alpha_4\rangle$, $\langle d_2\rangle $ and $\langle d_3\rangle$. An explicit MD verification of this structural transition is given in Figs.~\ref{fig:fig4} to \ref{fig:fig7}.}\label{fig:fig3}
\end{figure}

\section{Increasing the transition temperature with tensile strain}\label{sec:5}
MD calculations uncovering phase transitions on group-IV monochalcogenide monolayers that are based on a NPT ensemble\cite{nl,Mehrshad} agree with experimental observation concerning the collapse of $\langle\Delta \alpha\rangle$\cite{KaiSCIENCE} and display a delicate correspondence with classic theoretical results on phase transitions in 2D lattices.\cite{Potts} It will now be shown how uniaxial strain permits raising $T_c$ up to the large values reported in Model 2, where lattice parameters are not allowed to evolve with temperature.

To this end, and as illustrated in Fig.~\ref{fig:fig3}(a), SnSe monolayers were subjected to a one or two percent uniaxial tensile strain along the direction defined by either $\mathbf{a}_1$ or $\mathbf{a}_2$ at zero temperature and relaxed, still at zero temperature, afterwards. Lattice parameters prior and after the structural optimization are reported in Table \ref{ta:ta3}.

Uniaxial strain impedes the creation of a square structure at $T_c$, and $\langle\Delta \alpha\rangle$ remains non-zero  through the transition, as displayed in Fig.~\ref{fig:fig3}(b). The introduction of this symmetry-breaking constraint\cite{Kadanoff} reduces the original four-fold degeneracy onto a two-fold one.

As highlighted in Fig.~\ref{fig:fig3}(c) there are three dissimilar angles prior to the transition, and two dissimilar ones once the transition takes place. Similarly, as indicated in Fig.~\ref{fig:fig3}(d), $\langle d_2\rangle$ and $\langle d_3\rangle$ become equal at $T_c$. This happens as the tilt $\langle\delta\rangle$ and $\langle\theta\rangle$ both turn to zero, thus quenching the in-plane electric dipole too.

\begin{table}[tb]
 \caption{Lattice parameters of strained SnSe at zero temperature prior ($a_{1,0}$, $a_{2,0}$) and after ($a_{1}$, $a_{2}$) a structural optimization. Here, $\epsilon=\delta a_1/a_1$, or $\epsilon=\delta a_2/a_2$, accordingly.}\label{ta:ta3}
\centering
\begin{tabular}{|c||c|c|c||c|c|c|}
\hline
$\epsilon$ & $a_{1,0}$ (\AA) & $a_{2,0}$ (\AA) & $\frac{a_{1,0}}{a_{2,0}}$ & $a_1$ (\AA)& $a_2$ (\AA)& 	$\frac{a_1}{a_2}$\\
\hline
\hline
0.01 ($a_1$) &4.5160 & 4.3264 & 1.044 & 4.5160 & 4.3200 &  1.045\\
0.02 ($a_1$) &4.5600 & 4.3264 & 1.054 & 4.5600 & 4.3020 &  1.060\\
\hline
0.01 ($a_2$) &4.4873 & 4.3750 & 1.026 & 4.4400 & 4.3750 &  1.015\\
0.02 ($a_2$) &4.4873 & 4.4191 & 1.015 & 4.3551 & 4.4191 &  0.986\\
\hline
\hline
\end{tabular}
\end{table}

The structural transition described in Fig.~\ref{fig:fig3} is the one argued for in Ref.~\cite{Yang}, where only two degenerate ground states exist. It will be explicitly verified through MD calculations on uniaxially-strained samples in Figs.~\ref{fig:fig4} to \ref{fig:fig7}, that display the configurational energy $\langle U\rangle$, the electric dipole, and structural order parameters that include $\langle \theta \rangle$, $\langle \delta \rangle$, lattice parameters $\langle a_1\rangle$ and $\langle a_2\rangle$, as well as the parameters $\langle \Delta \alpha\rangle$, angles and distances that were highlighted in Figs.~\ref{fig:fig3}(b-d).

Similar to previous studies on non-strained samples, an 8$\times$8 supercell is built out of the strained unit cells at zero temperature afterwards, and the MD simulation box is kept fixed along the strained direction throughout the thermal evolution, by an in-house modification of the computational tool. MD calculations on the NPT ensemble ran for over 30,000 femtoseconds at selected temperatures.

Figure~\ref{fig:fig4} displays a 2D structural phase transition of a SnSe monolayer under 1\% tensile uniaxial strain along $a_1$ that is captured in Fig.~\ref{fig:fig4}(a) by a sudden increase of $\langle U\rangle$ at a $T_c=390$ K that is higher than its $T_c=175$ K value in Fig.~\ref{fig:fig2}(a) and is a result of the structural constraint. The saturation value of $\langle U\rangle$ is also larger than that seen in Fig.~\ref{fig:fig2}(a).

The order parameters $\langle \theta\rangle$, $\langle\delta\rangle$ and the electric dipole $\langle p_x\rangle$ show an identical dependence on temperature in Figs.~\ref{fig:fig4}(b-d). These identical trends can be understood from the fact that $\langle\delta\rangle$ is the in-plane separation among the positive group-IV element and the negative chalcogen (e.g., atoms $\mathbf{b}_2$ and $\mathbf{b}_3$), that turns the in-plane electric dipole $\langle p_x\rangle$ on, while $\langle\theta\rangle$ is linearly proportional to $\langle \delta\rangle$ for small angles.

The lattice parameter $\langle a_1\rangle$ in Fig.~\ref{fig:fig4}(e) can be obtained either from the fixed length of the constrained supercell, or from the distance among identical basis atoms belonging to consecutive unit cells. The second choice, displayed in Figs.~\ref{fig:fig4} through \ref{fig:fig7}, permits adding information about out-of-plane oscillations at finite temperature and confers $\langle a_1 \rangle$ with a slight slope and an error bar.

The orthogonal and unconstrained lattice vector $\mathbf{a}_2$ increases its magnitude with temperature due to a positive coefficient of thermal expansion. Nevertheless, $\langle \Delta\alpha\rangle$ in Fig.~\ref{fig:fig4}(f) remains non-zero through this transition:  {\em $\langle\Delta\alpha\rangle$ is not a good measure for the structural transition of strained samples.}

Despite of the lack of converging values of $\langle \Delta\alpha\rangle$ to 0 in Fig.~\ref{fig:fig4}(e), Figs.~\ref{fig:fig4}(g) and \ref{fig:fig4}(h) show a convergence of $\langle\alpha_1\rangle$ onto  $\langle\alpha_3\rangle$ at $T_c$ that is similar to the one shown in Fig.~\ref{fig:fig3}(c). Similarly, $\langle d_2\rangle=\langle d_3\rangle$ at $T_c$ in Fig.~\ref{fig:fig4}(h), which is consistent with the transition depicted in Fig.~\ref{fig:fig3}(d).

Given that $\langle p_x\rangle$ is quenched in Fig.~\ref{fig:fig4}(d), the transition of a group-IV monolayer under uniaxial tensile strain bears resemblance to a transition on a NVT ensemble;\cite{Yang} the exception being the release of $a_2$ to vary, a condition consistent with a SnSe monolayer clamped at two opposite ends only.

Figure \ref{fig:fig5} shows the structural transition when the strain is raised to a still small value of 2\%. The transition is similar to the one described in Fig.~\ref{fig:fig4}, but now $\langle U\rangle$ doubles it value when compared to its magnitude in Fig.~\ref{fig:fig4} while $T_c$ continues to increase, thus demonstrating the high degree of tunability of $T_c$ with moderate tensile strain.

Compressive strain is hard to achieve in 2D materials, but as seen in Fig.~\ref{fig:fig6}, tensile strained can also be applied along the short lattice vector $a_2$, thus favoring a square structure. The larger magnitude of $T_c$ in Fig.~\ref{fig:fig6} indicates that any constraint on the original four-fold degenerate structure increases $T_c$. As discussed before,\cite{nl} an unstrained unit cell requires two ``turning events'' to switch its polarization by 180 degree: a direct flip of polarization from $A_{\rightarrow}$ to $A_{\leftarrow}$ requires an energy of $2J$, while a two-step flip (either $A_{\rightarrow}$ to $B_{\uparrow}$ to $A_{\leftarrow}$, or $A_{\rightarrow}$ to $B_{\downarrow}$ to $A_{\leftarrow}$) only requires overcoming a barrier equal to $J$ at each 90-degree flip. In favoring a pair of degenerate ground states over the other two, one reduces the probability of the two-step transition to favor a transition through the larger ($2J$) barrier, hence raising $T_c$.

Fig.~\ref{fig:fig6} is similar to Figs.~\ref{fig:fig4} and \ref{fig:fig5}, but Fig.~\ref{fig:fig6}(e) shows a decrease of $\langle a_1\rangle$ towards $\langle a_2\rangle$ that is suddenly suppressed at higher temperature. The sudden change of $\langle a_1\rangle$ and $\langle a_2\rangle$ with temperature is due to the thermal softening of elastic constants in these 2D materials.

\section{Setting the orientation of the electric dipole with tensile strain}\label{sec:6}

\begin{figure}[tb]
\includegraphics[width=0.48\textwidth]{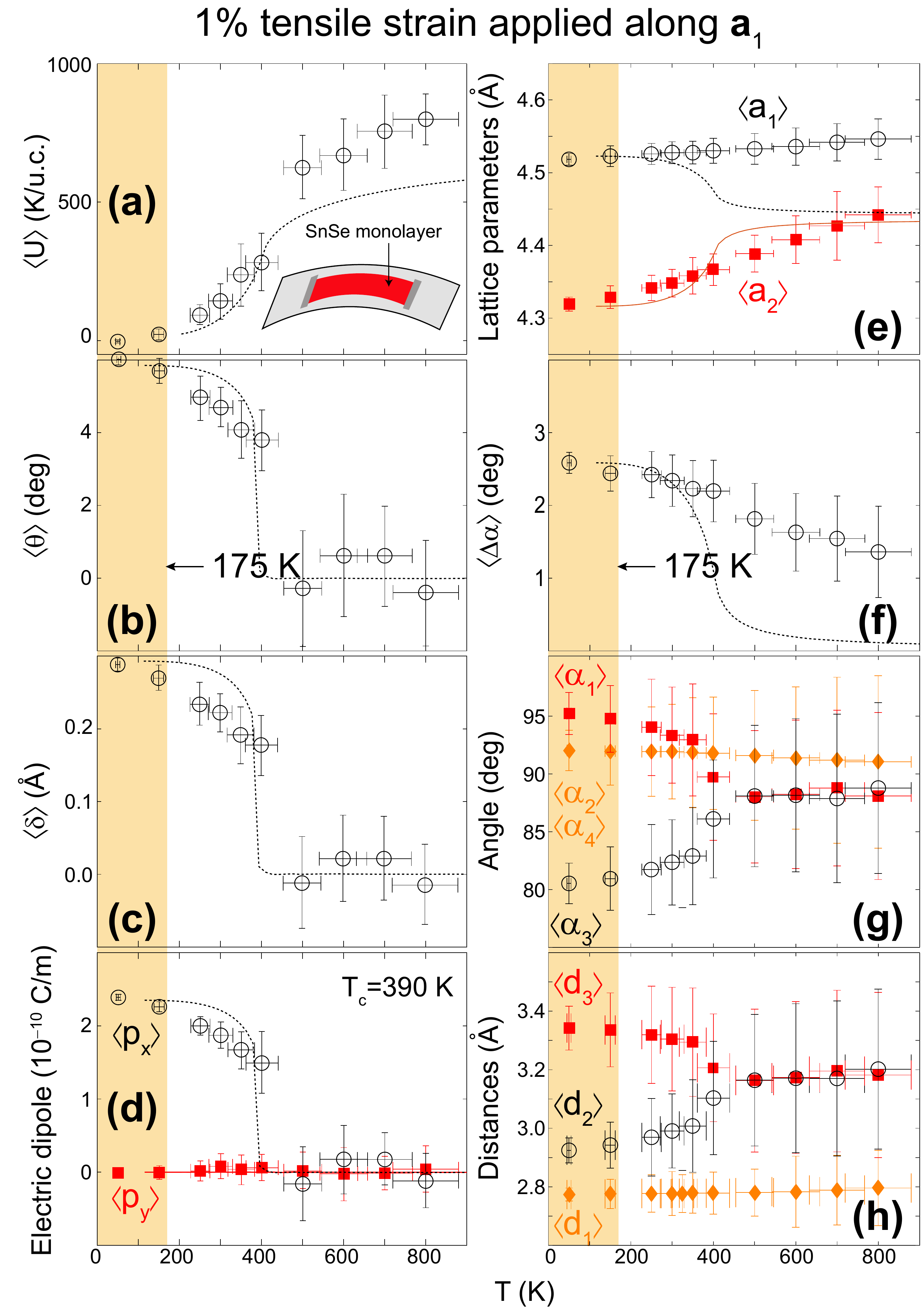}
\caption{Structural transition of a SnSe monolayer under an initial 1\% uniaxial tensile strain along $a_1$: while $T_c$ is signalled by the sudden agreement of lattice parameters ($\langle a_1\rangle =\langle a_2\rangle$) and the collapse of $\langle \Delta\alpha \rangle$ on unstrained samples, a strained sample preserves a rectangular shape. Nevertheless, its intrinsic dipole turns to zero as in-plane angles and distances take on two values for $T\ge T_c$, instead of three for $T< T_c$. The increase on $T_c$ with respect to the value in an unstrained sample ($175$ K) is emphasized by the yellow rectangle. Fitting curves are thermodynamical averages arising from Eqn.~\ref{eq:model}.}\label{fig:fig4}
\end{figure}

\begin{figure}[tb]
\includegraphics[width=0.48\textwidth]{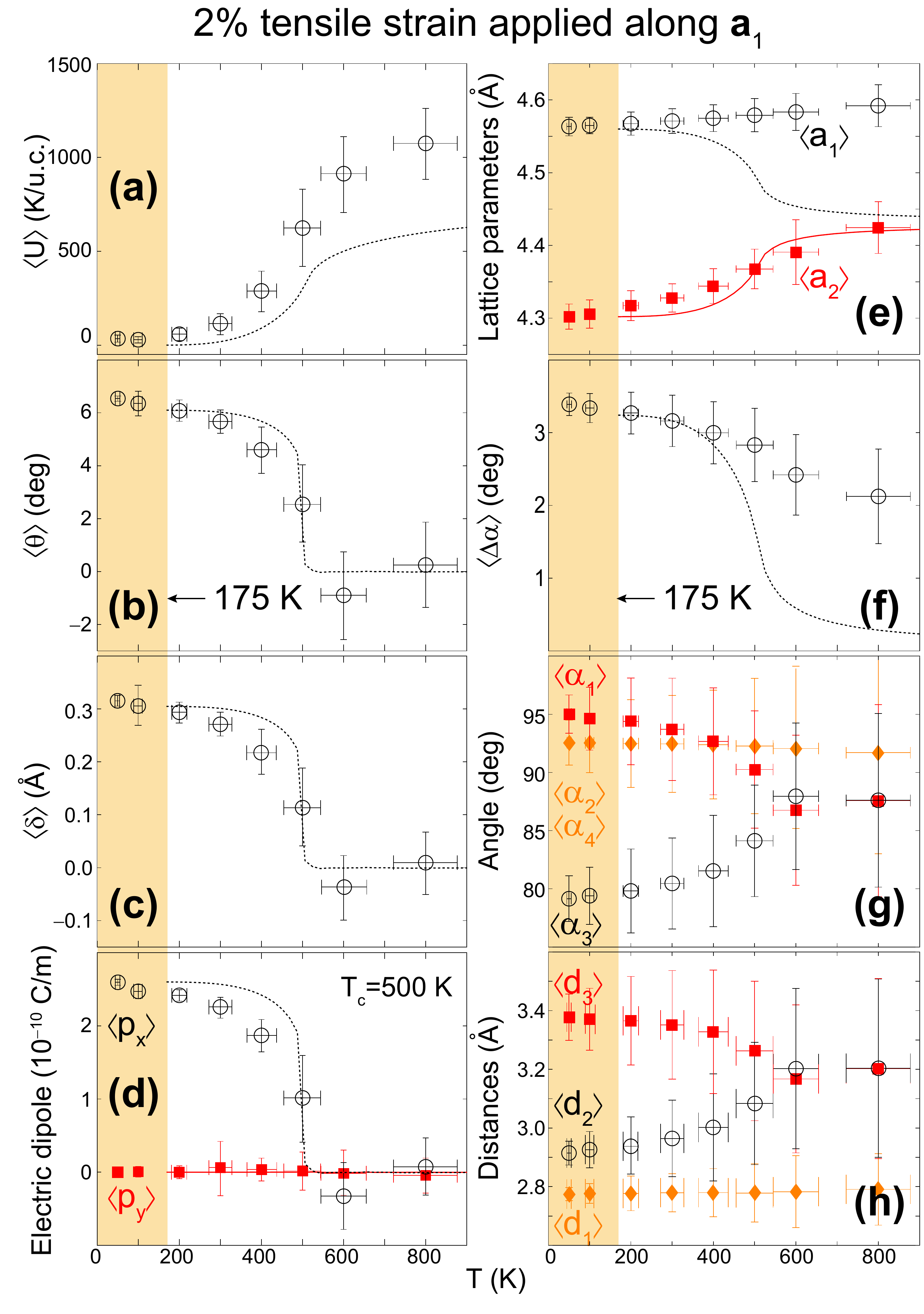}
\caption{Structural transition of a SnSe monolayer under an initial 2\% uniaxial tensile strain along $a_1$, that raises $T_c$ to 470 K. Fitting curves are thermodynamical averages arising from Eqn.~\ref{eq:model}.}\label{fig:fig5}
\end{figure}

$\langle a_1\rangle$ is larger than $\langle a_2\rangle$ in Figs.~\ref{fig:fig4} through \ref{fig:fig6}. Nevertheless, the SnSe monolayer aligns its in-plane dipole under a threshold uniaxial tensile strain along $a_2$, to become parallel to the direction of the external uniaxial tensile strain: uniaxial tensile strain can be used to orient the direction of the intrinsic in-plane electric field.

As shown in Table \ref{ta:ta3} and Fig.~\ref{fig:fig7}, a 2\% strain along the initially smaller in-plane lattice vector $a_2$ is sufficient to make $a_2$ larger than $a_1$, and MD calculations indicate that the electric dipole realigns to be parallel to the $y-$direction: in Fig.~\ref{fig:fig7}, angles $\langle \alpha_2\rangle$ and $\langle \alpha_4\rangle$ take on dissimilar values at zero temperature, and converge at $T_c$, while $\langle \alpha_1\rangle$ and $\langle \alpha_3\rangle$ remain identical through the transition. In contrast, Figs.~\ref{fig:fig4}(e), \ref{fig:fig5}(e) and \ref{fig:fig6}(e) display different magnitudes of $\langle \alpha_1\rangle$ and $\langle \alpha_3\rangle$ at zero temperature that converge at $T_c$, while $\langle \alpha_2\rangle$ and $\langle \alpha_4\rangle$ remain identical, while the in-plane electric dipole was oriented along the $x-$direction. $T_c$ is raised to 250 K in this scenario.

\section{Phase transition of strained monolayers in a two-parameter model}\label{sec:7}

As seen in Fig.~\ref{fig:fig2}, a clock model with one single fitting parameter $J$ is sufficient to understand the phenomenology of unstrained group-IV monochalcogenides. In order to emphasize the basic physical behavior over numerical details, we wish to maintain the simplicity of that model in describing strained monolayers.

\begin{figure}[tb]
\includegraphics[width=0.48\textwidth]{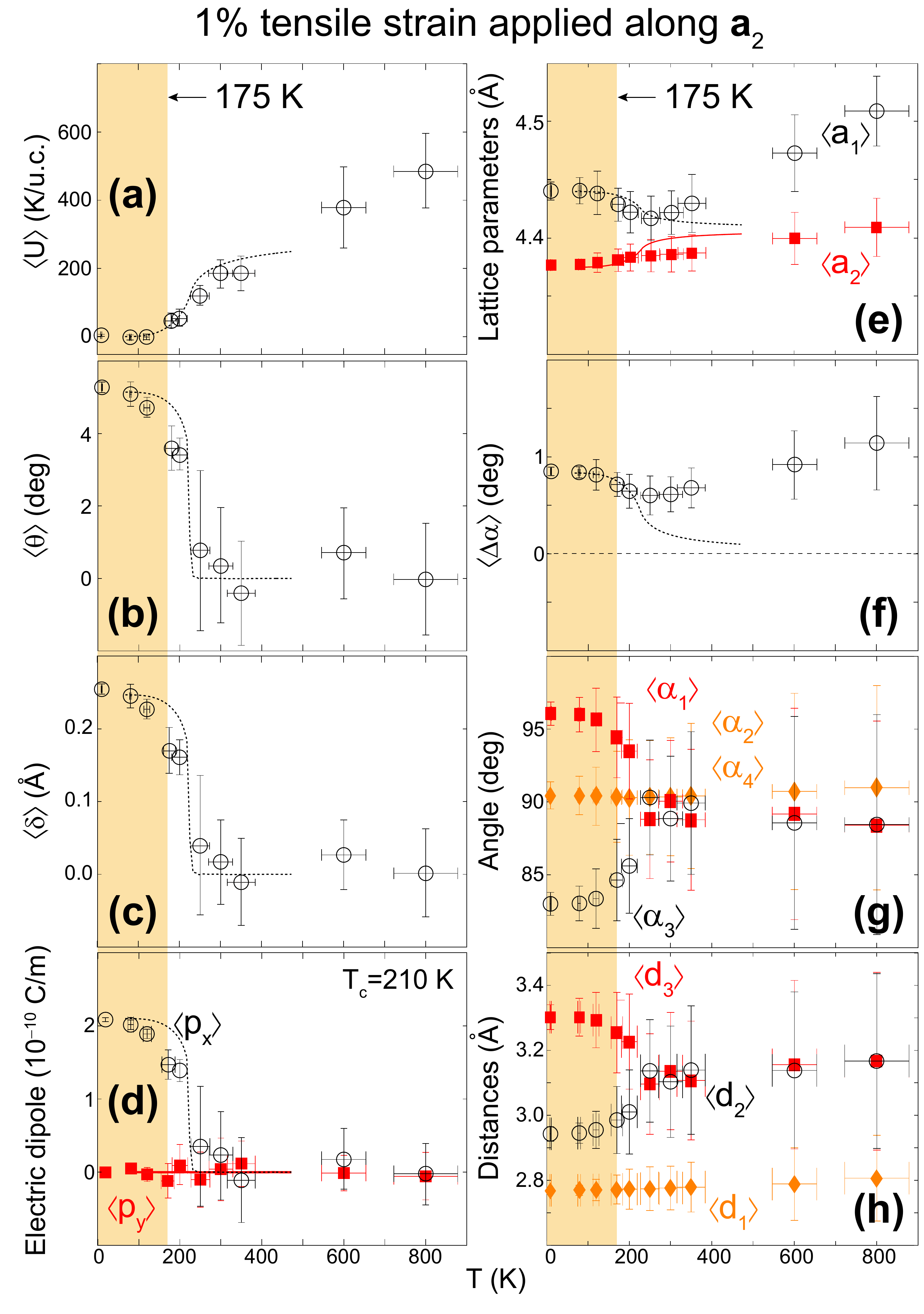}
\caption{SnSe monolayer under 1\% uniaxial strain along $a_2$: given that the transition still occurs outside of the yellow-marked areas, $T_c$ still increases when the shorter lattice parameter $a_2$ was elongated by 1\%.  Fitting curves are thermodynamical averages arising from Eqn.~\ref{eq:model}.}\label{fig:fig6}
\end{figure}

\begin{figure}[tb]
\includegraphics[width=0.48\textwidth]{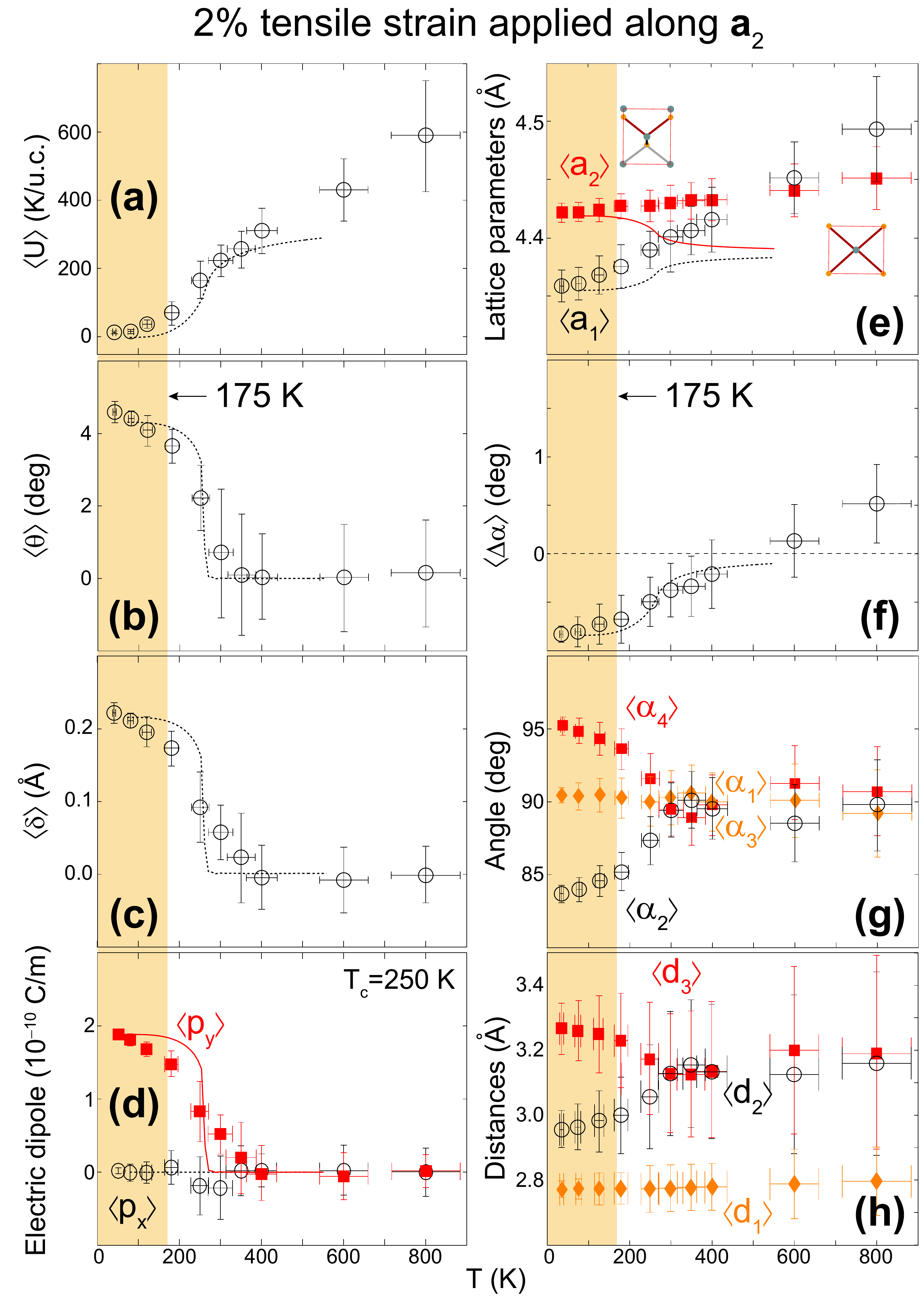}
\caption{SnSe monolayer under 2\% uniaxial strain along $a_2$: as the transition occurs for $T_c$ outside the yellow box, $T_c$ still increases when the shorter lattice parameter is elongated by 2\%, the orientation of the electric dipole flips in order to point along the longest lattice vector, making $\langle \delta\rangle$ align along the $y-$direction in subplot (c), so that $\langle p_y\rangle $ is non-zero in subplot (d). (e) $\langle a_1\rangle$ becomes larger than $\langle a_2\rangle$ again at a temperature larger than $T_c$, making $\langle\Delta \alpha\rangle$ in subplot (f) change sign. The dipole orientation along the $y-$axis comes about from the angles in subplot (g) that are different before $T_c$, when compared with those in Figs.~\ref{fig:fig4} to \ref{fig:fig6}.  Fitting curves are thermodynamical averages arising from Eqn.~\ref{eq:model}.}\label{fig:fig7}
\end{figure}

As indicated in Section \ref{sec:4}, the relation among $T_c$ and $J$ increases by decreasing the number of degenerate ground states. This observation implies that the increase of $T_c$ observed in Figs.~\ref{fig:fig4} to \ref{fig:fig7} with respect to its magnitude on an unstrained sample, could in principle be assigned to the favoring of two structural ground states ({\em i.e.}, those two parallel to the applied strain) and makes it more energy costly to occupy the two states that are parallel to the direction of the applied strain. An interaction of the form $-|\mathbf{p}_i(\mathcal{\epsilon})\cdot\mathcal{\epsilon}|$ enforces the preference of two degenerate states over the other two, and sets the system in between a Potts model with four degenerate ground states when previous term is turned off, and an Ising model when this term is on and set larger than $J$, such that $T_c\sim [1.1(4)--2.2(7)]J$ depending on the magnitude of strain. Previous statements imply that strain lowers the initial symmetry of the structure and lowers the number of degenerate ground states.\cite{Kadanoff}. The preference of two states over the other two implies that a square structure is not found at $T_c$ as well, such that $|\Delta \alpha|> 0$ at $T_c$.

This way, the effective dynamics of strained samples takes the following form:
\begin{equation}\label{eq:model}
U=-J\sum_i\left(1-\sum_{\langle i,j\rangle}\cos(\Theta_i-\Theta_j)\right)-h\sum_i|\mathbf{p}_i(\mathcal{\epsilon})\cdot\mathcal{\epsilon}|,
\end{equation}
where $i$ runs over $n-$individual sites, $\langle i,j\rangle$ implies a sum over next-nearest neighbors, and $\Theta_i$ is the (discrete) dipole orientation, which can take on four values that correspond to the four degenerate ground states on the unstrained sample.

The first term to the right of Eqn.~\eqref{eq:model} is similar to the one given in Ref.~\cite{nl}. As discussed earlier, the second term reduces the original four-fold degeneracy because it favors orientations of the electric dipole that are parallel to the direction of the applied strain, turning the system into an Ising (two-fold degenerate) lattice, and hence yielding $T_c$ in between 1.14$J$ when the first term dominates and 2.27$J$ when the second term does. We consider strain $\epsilon$ parallel to either $\mathbf{a}_1$ or $\mathbf{a}_2$ and dipole moments pointing parallel or anti-parallel to the lattice vectors.

The dynamics expressed by Eqn.~\eqref{eq:model} were employed in an in-house Monte Carlo solver on a 60$\times$60 supercell, and the solid trendlines in Figs.~\ref{fig:fig4} through \ref{fig:fig7} are results from the model that fully describe the MD phenomenology. In order for the model to best describe MD data, we found it necessary to increase the magnitude of $J$. This is, strain sets a preference for two degenerate ground states, but it also increases the elastic energy barrier. The parameters employed in obtaining the dashed curves in Figs.~\ref{fig:fig4} through \ref{fig:fig7} are listed in Table \ref{ta:ta4}.

$\langle U\rangle$ is the expectation value of $U$ given in Eqn.~\eqref{eq:model}, and writing the probability of a given dipole orientation as $\langle \rightarrow\rangle$, $\langle \leftarrow\rangle$, $\langle \uparrow\rangle$ and $\langle \downarrow\rangle$, which are all functions of temperature, condition 2 in Section 1 is established by setting  $\langle \rightarrow\rangle=1$ at zero temperature. This way, the lattice parameters and other order parameters are estimated by:
\begin{eqnarray*}
\langle a_1(\boldsymbol{\epsilon})\rangle=\\
\frac{a_1(T=0,\boldsymbol{\epsilon})(\langle\rightarrow\rangle+\langle\leftarrow\rangle)+a_2(T=0,\boldsymbol{\epsilon})(\langle\uparrow\rangle
+\langle\downarrow\rangle)}{\langle\rightarrow\rangle+\langle\leftarrow\rangle+\langle\uparrow\rangle+\langle\downarrow\rangle},
\end{eqnarray*}
\begin{eqnarray*}
\langle a_2(\boldsymbol{\epsilon})\rangle=\\
\frac{a_1(T=0,\boldsymbol{\epsilon})(\langle\uparrow\rangle+\langle\downarrow\rangle)+a_2(T=0,\boldsymbol{\epsilon})(\langle\rightarrow\rangle
+\langle\leftarrow\rangle)}{\langle\rightarrow\rangle+\langle\leftarrow\rangle+\langle\uparrow\rangle+\langle\downarrow\rangle},
\end{eqnarray*}
$$
\langle \Delta \alpha(\boldsymbol{\epsilon})\rangle=\frac{\langle a_1(\boldsymbol{\epsilon})\rangle}
             {\langle a_2(\boldsymbol{\epsilon})\rangle}-1,
$$
$$
\langle p_x(\boldsymbol{\epsilon})\rangle=\frac{p_x(T=0,\boldsymbol{\epsilon})(\langle\rightarrow\rangle-\langle\leftarrow\rangle)}
{\langle\rightarrow\rangle+\langle\leftarrow\rangle+\langle\uparrow\rangle+\langle\downarrow\rangle},
$$
and
$$
\langle p_y(\boldsymbol{\epsilon})\rangle=\frac{p_x(T=0,\boldsymbol{\epsilon})(\langle\uparrow\rangle-\langle\downarrow\rangle)}
{\langle\rightarrow\rangle+\langle\leftarrow\rangle+\langle\uparrow\rangle+\langle\downarrow\rangle},
$$
and shown by black dashed and red solid lines in Figs.~\ref{fig:fig2} (for $\boldsymbol{\epsilon}=\mathbf{0}$, $J=150$ K, and $h=0$), and \ref{fig:fig4} to \ref{fig:fig7}. There, $\langle \theta\rangle$ and $\langle \delta\rangle$ are proportional to $\langle \mathbf{p}\rangle$, and the zero-temperature values are taken from Table \ref{ta:ta3}. The qualitative agreement among the full-scale MD data and the results from the model stands out given the simplicity of the latter: though the model could be improved, it captures the essential effects of strain on structure.

\begin{table}[tb]
 \caption{Magnitudes of model parameters and $T_c$.}\label{ta:ta4}
\centering
\begin{tabular}{|c||c|c|c|c|c||c|}
\hline
$\epsilon$ & $J$ (K) & $h$   & $p_x$ & $\delta$ (\AA)& 	$\theta$ (deg) & $T_c$ (K)\\
 &  & ($10^{10}\frac{Km}{C}$) & ($10^{-10}\frac{C}{m}$) & & & 	\\
\hline
\hline
0.01 ($a_1$) &330 & 25.53   & 2.35 & 0.295 &  5.9& 390\\
0.02 ($a_1$) &375 & 38.46  & 2.60 & 0.305 &  6.1& 500\\
\hline
0.01 ($a_2$) &175 & 5.00   & 2.00 & 0.270 &  5.4& 210\\
0.02 ($a_2$) &175 & 51.28  & 1.95 & 0.215 &  4.3& 250\\
\hline
\hline
\end{tabular}
\end{table}

\section{Conclusions}

To conclude, this manuscript improves the present understanding of two-dimensional structural phase transitions on two-dimensional materials beyond graphene.

The conditions for 2D structural transitions are: the existence of degeneracies on the ground state unit cell, a path among degenerate ground states that has an energy barrier smaller than the melting point, and the existence of sufficiently large monodomains displaying a given ground state.

Unstrained group-IV monochalcogenide monolayers possess four switchable ground states. These materials undergo a 2D structural transition at finite temperature provided the lattice parameters evolve freely and unconstrained, such that all four ground states are sampled. Sampling of the four ground states is essential for theory to describe experimentally-observed transitions that are triggered by the collapse of $\langle\Delta \alpha\rangle$ to zero.

Constraining the unit cell lattice vectors to their magnitude at zero temperature while discussing finite-temperature properties amounts to applying strain and it raises the transition temperature from its magnitude on an {\em intrinsic}, unstrained sample.

The transition temperature, and even the orientation of the in-plane intrinsic electric dipole can be widely controlled by moderate uniaxial tensile strain. These MD results can be qualitatively cast onto an extension of the clock model.

The results from the present study will assist in establishing a solid theoretical background for further work in phase transitions in two-dimensional materials, and their effects on material properties, and offer intriguing connections among topics in soft-condensed matter and novel two-dimensional atomic materials.

\section{Methods}\label{sec:methods}
The energy landscape and the dependency of order parameters of SnSe monolayers on $r$ in Fig.~\ref{fig:fig1} were obtained with the {\em SIESTA} DFT code\cite{siesta} in calculations carried out on the unit cell at zero temperature and with van der Waals corrections within the consistent-exchange vdW-DF-cx functional\cite{BH}. The pseudopotentials with van der Waals corrections have cutoff radii as listed for PBE pseudos in Ref.~\cite{Pablo}. Calculations proceeded with a 18$\times$18 $k-$point grid, and a mesh cutoff of 300 Ry for the Poisson solver was employed as well. The mesh from which Fig.~\ref{fig:fig1}(a) was drawn included 50 independent values of $a_1$ and $a_2$. As indicated in the main text, the unit cells along the $a_1=a_2$ line were obtained on structures that have an explicit zero net in-plane dipole.

The structures listed in Tables \ref{ta:ta1} and \ref{ta:ta2} were obtained with the {\em SIESTA}, {\em VASP} \cite{VASP}, and {\em Quantum Espresso} \cite{QE} computer codes, as listed.

All results obtained within plane-wave, pseudopotential density-functional theory methods (e.g., {\em VASP} and {\em Quantum Espresso}) employ projector-augmented wave \cite{blochl} pseudopotentials that are tuned against the  open-source pseudopotential library.\cite{vasppseudos,GBRV}.

The calculations within {\em VASP} employ a $15\times15$ $k-$point grid and a cutoff energy of 37 Ry. The force convergence criteria was set to $10^{-3}$ eV/\AA. In {\em Quantum Espresso} calculations a $15\times15$ $k-$point grid was also employed, with cutoff energy of 40 Ry, and a force convergence criteria of $10^{-4}$ eV/\AA. van der Waals corrections in the {\em VASP} code were turned on by employing the following flags: $GGA = OR$; $LUSE\_VDW = .TRUE.$, and $AGGAC = 0.0000$. Espresso calculations with van der Waals corrections employed the $vdW-DF-obk8$ flag.

The landscape shown in Fig.~\ref{fig:fig1}(a) is extremely flat near the local minima for a regular force minimization process with standard limits (e.g., a force tolerance of 0.001 eV/AA) to reach the lowest-energy configuration. For this reason, a meshing of $a_1$ and $a_2$ around the minimum-energy structures was employed to truly guarantee that the absolute minima had been reached. This should help reduce the current spread in known structural estimates.

Figure~\ref{fig:fig2} re-expresses results from previous calculations\cite{Mehrshad} in the language of Refs.~\cite{KaiSCIENCE} and \cite{Yang}. These results arise from {\em ab initio} MD calculations with the {\em SIESTA} code that were performed for up to 30,000 fs on the NPT ensemble, with basis sets and input parameters similar to those listed in previous paragraph for consistency.

 Table \ref{ta:ta2} lists structural parameters for a SnSe monolayer with the {\em VASP} code within the PBE\cite{PBE} approximation for exchange-correlation. Here we restate the existence of a systematic underestimation of the ratio $a_1/a_2$ in DFT calculations that can be expressed as follows:
$$1\le (a_1/a_2)_{LDA} < (a_1/a_2)_{PBE} < (a_1/a_2)_{vdW}.$$

The results in Figs.~\ref{fig:fig4} through \ref{fig:fig7} where obtained with the {\em SIESTA} code using input parameters that are similar to those listed two paragraphs above. In the NPT ensemble, pressure induces a force that pushes the periodic walls constraining the 2D material. Here, pressure is overwritten to zero along the direction constrained by the application of uniaxial strain, which effectively fixes the wall along that constrained direction.

Numerical limitations in the theoretical understanding of group-IV monochalcogenides must also be properly acknowledged in order to foresee opportunities for further work. For example, the magnitude of $J$ could be contrasted against other van der Waals implementations,\cite{cooper1,cooper2} other approaches like Quantum Montecarlo,\cite{Shulen} and available experiments.\cite{KaiSCIENCE} Similar to status of bulk ferroelectrics, experimental and theory-based transition temperatures tend not to be in perfect agreement, which does not preclude a complete theoretical description of the fundamental physical picture at hand.


This work was funded by an Early Career Grant from the DOE (DE-SC0016139; S.B.L. and T.P.K). Calculations were performed at Arkansas High Performance Computing Center's {\em Trestles}, which is funded through multiple National Science Foundation grants and the Arkansas Economic Development Commission. Conversations with Kai Chang are gratefully acknowledged.



%

\end{document}